# Building Reality Checks into the Translational Pathway for Diagnostic and Prognostic Models

## Authors

Lendrem DW[1,2], Lendrem BC[1,3,4], Pratt AG[1,2,3], Tarn JR[1], Skelton A1[1,5], James K[6,7], McMeekin P[8,9], Linsley M[10], Gillespie C[10], Cordell H[11], Ng WF[1,2,3], Isaacs JD[1,2,3]

[1] Institute of Cellular Medicine, Newcastle University, Newcastle upon Tyne, UK
[2] NIHR Newcastle Biomedical Research Centre, Newcastle University, Newcastle upon Tyne, UK
[3] Newcastle upon Tyne Hospitals NHS Trust, Newcastle upon Tyne, UK
[4] NIHR Newcastle In Vitro Diagnostics Co-operative, Newcastle University, Newcastle upon Tyne, UK
[5] Bioinformatics Support Unit, Newcastle University, Newcastle upon Tyne, UK
[6] Interdisciplinary Computing & Complex Biosystems Research Group, Newcastle University, Newcastle upon Tyne, UK
[7] Centre for Bacterial Cell Biology, Newcastle University, Newcastle upon Tyne, UK
[8] Institute of Health & Society, Newcastle University, Newcastle upon Tyne, UK
[9] School of Health Sciences, Northumbria University, Newcastle upon Tyne, UK
[10] Department of Mathematics & Statistics, Newcastle University, Newcastle upon Tyne, UK
[11] Institute of Genetic Medicine, Newcastle University, Newcastle University, Newcastle upon Tyne, UK

# Building Reality Checks into the Translational Pathway for Diagnostic and Prognostic Models

## Abstract

There has been a significant increase in the number of diagnostic and prognostic models published in the last decade.  Testing such models in an independent, external validation cohort gives some assurance the model will transfer to a naturalistic, healthcare setting.  Of 2,147 published models in the PubMed database, we found just 120 included some kind of separate external validation cohort.  Of these studies not all were sufficiently well documented to allow a judgement about whether that model was likely to transfer to other centres, with other patients, treated by other clinicians, using data scored or analysed by other laboratories.  We offer a solution to better characterizing the validation cohort and identify the key steps on the translational pathway for diagnostic and prognostic models.

# Building Reality Checks into the Translational Pathway for Diagnostic and Prognostic Models

## Background

The last decade has seen a marked increase in diagnostic and prognostic prediction models in medicine. The growth of interest in biomarkers, personalized and stratified medicine, coupled with the rise of machine learning and statistical algorithms aiding the building of such diagnostic and prognostic tools[1-3] has led to a proliferation of prediction models in recent years - see Supplementary Figure 1. However, the quality and reporting of studies to support such prediction models is often poor[4] prompting the development of reporting guidelines for studies of observational data[5], tumour markers[6], diagnostic accuracy[7], genetic risk prediction[3], and multivariable prediction models[4]– the STROBE, REMARK, STARD, GRIPS and TRIPOD guidelines respectively. In particular, the TRIPOD guideline offers specific recommendations around studies developing or validating prediction models[4].

As a result, most studies now use some kind of validation cohort to evaluate the model. Usually some portion of the data is retained for testing models. Internal validation tools – such as split-sampling, cross validation, and bootstrapping - are an essential part of the model building process. Internal validation directly addresses sampling variability: minimizing the risk of over-fitting models; minimizing the impact of outliers or other influential data; and guarding against chance partitions of the data impacting upon the model. Unfortunately, they tell us little or nothing about variation in the study population or clinical setting. They tend to be overoptimistic giving good "apparent" performance while giving little or no assurance on how the prediction model is likely to perform in real life[8]. Internal validation is a necessary, but not sufficient, condition to ensure the robust transfer of diagnostic and prognostic tools. However, in medicine it is critically important to know if the model generalizes to other patients, assessed and treated by other clinicians, at other sites, using assays performed in other laboratories.

## Main Text

This problem is essentially one of technology transfer or 'transportability'[9]. And internal validation methods tell us little or nothing about how the technology – the prediction model – performs when transferred to another setting. For this reason, the performance of a model must be tested using data other than that used for the model development[10,11 12]. An independent, external validation cohort provides much stronger evidence that a prognostic or diagnostic model will perform in real life[13]. Without such a cohort, the "ecological validity" of the model may be severely limited. Bleeker *et al* [8] for example give an early paediatric infection example demonstrating how disappointing such prediction models can be when transferred to populations other than those in which they were initially developed. More recently, Sahami *et al* report on the limitations of prognostic models for proctocolectomy[14], and Damen *et al* [15] report on the limitations of cardiovascular risk models.

# Building Reality Checks into the Translational Pathway for Diagnostic and Prognostic Models

While external validation studies may be critical to understanding the performance of such models in real life, unfortunately such studies are relatively rare[16-18]. We searched the PubMed database for diagnostic and prognostic models published in medicine through to September 2016 (see Supplementary Table 1). The majority of studies employed some, often limited, form of internal validation. Of 2,147 published papers on prognostic or diagnostic modelling, 323 claimed to have tested, verified or validated the model using an independent OR external validation cohort – see Table 1. And of these studies, just 120 claimed to have tested, verified or validated the model using a separate independent AND external validation cohort. However, upon further review, not all of these 120 studies can be considered truly independent validation studies. Some confuse internal and external validation; others employ relatively 'soft' validation methods. For example, time-separated cohorts from the same centre, using the same protocols, examined and treated by the same clinicians, using test results from the same laboratory, may not capture all the sources of variability in a truly independent cohort. Even when models are validated using groups of patients from other centres, the laboratory analysis, diagnosis or scoring of results may be performed by the same team from a single centre – see Figure 1.

**Table 1:** *Validation of diagnostic and prognostic models in medicine for the period 1970-2016. Of 2,147 published models, just 120 included an external validation cohort.*

| Additional Search Terms | | prognostic model | diagnostic model | prognostic AND diagnostic |
|---|---|---|---|---|
| | | 1,584 | 563 | 2,147 |
| AND | validation OR test OR verification | 538 | 208 | 746 |
| AND | independent OR external | 288 | 35 | 323 |
| AND | external validation | 110 | 10 | 120 |

*Source: PubMed*

As part of the validation process it is essential to be clear what sources of variability - patient, clinician, clinical setting – are captured by the validation cohort. We need to know whether the validation cohort gives us any reassurance that a decision tool will transfer to other centres, other patients, other clinicians, in other clinical settings. Without such assurance it is most unlikely that model performance will be reproducible. To this end, we offer the Newcastle 'Validation Cohort' Checklist to assist reviewers in reality checking the validation and identifying those sources of variability captured by the validation cohort (Table 2). The goal is to provide greater assurance that a model will perform in a real world healthcare setting.

# Building Reality Checks into the Translational Pathway for Diagnostic and Prognostic Models

**Table 2:** *The Newcastle 'Validation Cohort' Checklist: the checklist captures some of the key sources of variability in a validation cohort for reality checking of diagnostic and prognostic models.*

| Key Questions | Reality Checks |
|---|---|
| The validation cohort offers some assurance that the model works: | |
| At different sites? | Geographical |
| At different times? | Temporal |
| Scored or diagnosed by different clinicians? | Clinical |
| Analyzed in different laboratories? | Laboratory |
| For different patient populations? | Population |

While diagnostic and prognostic models may be of some value even without such external validation, it is not unreasonable for funding agencies to expect clarity around the generalizability of those models, proposed validation plans and the prospective translational pathway. We urge researchers to be explicit about the sources of variability captured by the external validation cohort. The validation cohort needs to be sufficiently well documented that we can tell whether the model is likely to transfer to other centres, with other patients, treated by other clinicians, using data scored or analysed by other laboratories.

Of course, model validation and impact studies are but one step on the translational pathway. For our models to translate to real patient benefit, we need to build qualification and monitoring processes into that pathway – see Figure 2. Ultimately, to apply a model we need to know whether that model is valid in a given healthcare setting, healthcare trust or even an individual practice. Even an externally validated prognostic or diagnostic model is no more than a plausible candidate for routine use in the clinic. Without some kind of qualification study, it does not establish that the model is valid in an individual healthcare setting. Even after the validation and qualification of a prospective model, we need 'production' systems in place to monitor the performance of that model in a naturalistic healthcare setting. This permits models to be reviewed and recalibrated to reflect changes in the patient population, clinical protocols and training. The analogy we draw is that of the clinical trial process. The movement from development, through recalibration and validation, to qualification and monitoring parallels that of drug development: from early proof-of-concept studies, through dose finding to Phase 3 clinical trials, followed by post-marketing studies and surveillance. The emergence of national registries and access to 'big data' health records open up exciting possibilities for the external validation, recalibration and 'real-time, real-world' updating of diagnostic and prognostic tools[19]. Currently, it is questionable whether we have the national systems and data access permitting routine monitoring of such models. But that must be our goal.

**Building Reality Checks into the Translational Pathway for Diagnostic and Prognostic Models**

## Conclusions

While the internal validation cohort is an important part of the model building process, the external validation cohort gives greater assurance that a diagnostic or prognostic model will translate to the clinic. However, researchers need to be explicit about the sources of variability captured by the external validation cohort. The validation cohort needs to be sufficiently well documented that we can tell whether the model is likely to transfer to other centres, with other patients, treated by other clinicians, using data scored or analysed by other laboratories.

# Building Reality Checks into the Translational Pathway for Diagnostic and Prognostic Models

## Acknowledgements

AGP and JDI were supported by the National Institute for Health Research (NIHR) Newcastle Biomedical Research Centre based at Newcastle upon Tyne Hospitals NHS Foundation Trust and Newcastle University.  DWL was supported in part by the NIHR Biomedical Research Centre and Arthritis Research UK.

# Building Reality Checks into the Translational Pathway for Diagnostic and Prognostic Models

## Declarations

Authors' contributions: JDI, AGP, JRD, AS and KJ developed and discussed ideas around model validation. DWL and BCL drafted the manuscript. PM, ML, CG, HC, and FN reviewed and contributed to the final manuscript.

## Competing Interests

The authors declare that they have no competing interests.

## Ethics Approval and Consent to Participate

Not applicable.

## Consent for Publication

Not applicable.

## Availability of Data and Materials

The dataset used in this study is the PubMed database searchable through the NCBI website at https://www.ncbi.nlm.nih.gov/. Search terms used to generate Table 1 and Supplementary Figure 1 are given in Supplementary Table 1 – see Supplementary Data.

# Building Reality Checks into the Translational Pathway for Diagnostic and Prognostic Models

# Figures

*Figure 1: In order to assess the transferability of a diagnostic or prognostic model we need to be explicit about the sources of variability captured by our validation cohort. Imagine a model developed at **Centre 1** that was trained and tested on a population of patients **T**. This model might be expected to perform well on patients in the validation cohort $V_1$. However, these patients are derived from the same demographic region and treated using a shared clinical protocol. Evaluating performance at a separate centre – **Centre 2** – gives greater assurance that the model will travel to other centres in other regions. However, if this validation cohort includes some patients treated by the same physician, with test results performed at a shared laboratory, using images scored by the same radiographer then the performance of the model is unrealistically optimistic – the apparent performance is high. Excluding these individuals, **E**, from the validation cohort $V_2$ gives a separate, independent external validation cohort. This gives a more robust assessment of model performance.*

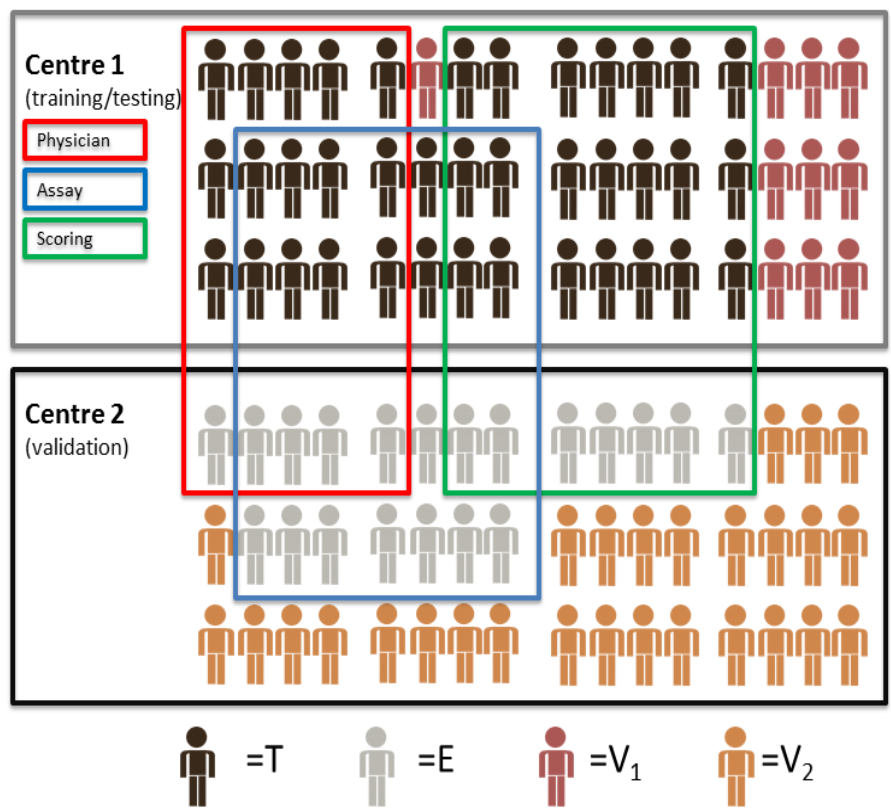

# Building Reality Checks into the Translational Pathway for Diagnostic and Prognostic Models

*Figure 2: For diagnostic and prognostic models to translate to real patient benefit we need a translational pathway characterised by 1) good internal validation during the modelling process, 2) well designed impact studies with well defined external validation cohorts during the validation process, 3) qualification studies to evaluate the tool in an individual setting, and 4) systems to monitor and feedback on model performance in real world healthcare settings.*

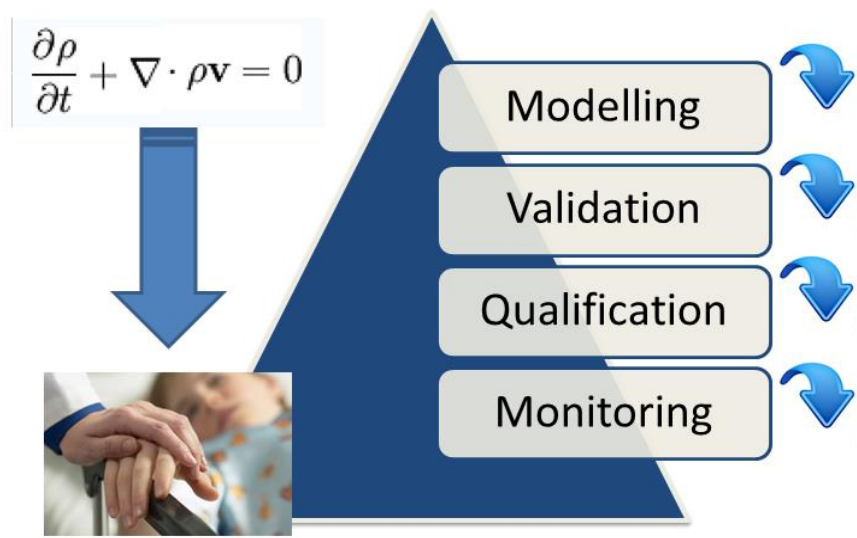

# Building Reality Checks into the Translational Pathway for Diagnostic and Prognostic Models

## Supplementary Data

**Supplementary Figure:**

**Supplementary Figure 1**: *Diagnostic and prognostic models published in medical journals between 1980-2015. The last ten years have seen a marked rise in published models reflecting the growing interest in biomarkers, stratified and personalized medicine, together with the development of machine learning algorithms and other statistical tools aiding the building of such models.*

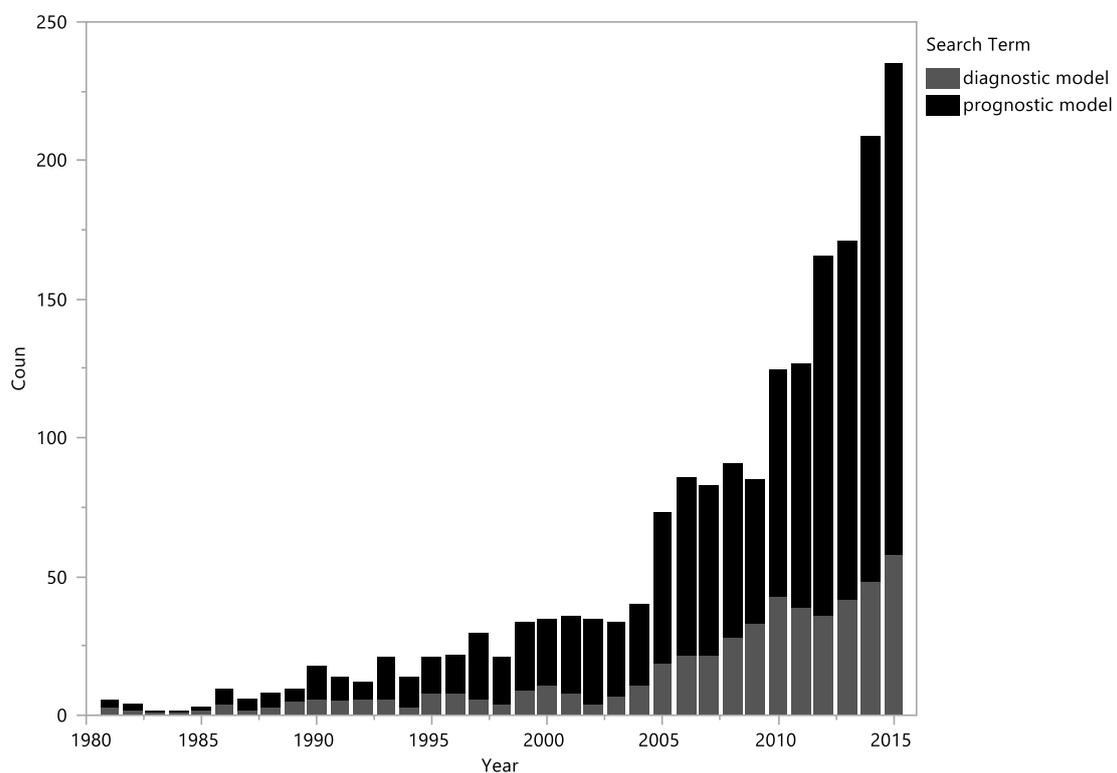

[Source: PubMed]

# Building Reality Checks into the Translational Pathway for Diagnostic and Prognostic Models

**Supplementary Table:**

**Supplementary Table 1**: *NCBI search terms and filters for diagnostic and prognostic models published in PubMed database medical journals through to 01/09/2016.*

*Source: NCBI*

| Models | PubMed Search Terms | Results |
|---|---|---|
| prognostic | Search "prognostic model"[Title/Abstract] Filters: Publication date to 2016/09/01 | 1,584 |
| | Search ((((prognostic model[Title/Abstract] AND ( "0001/01/01"[PDat] : "2016/09/01"[PDat] ))) AND ((validation[Title/Abstract]) OR (verification[Title/Abstract]) OR (test[Title/Abstract])) Filters: Publication date to 2016/09/01 | 538 |
| | Search ((((((((prognostic model[Title/Abstract] AND ( "0001/01/01"[PDat] : "2016/09/01"[PDat] ))) AND ((validation[Title/Abstract]) OR (verification[Title/Abstract]) OR (test[Title/Abstract]))) AND ( "0001/01/01"[PDat] : "2016/09/01"[PDat] ))) AND (independent[Title/Abstract] OR external[Title/Abstract] ) ( AND ( "0001/01/01"[PDat] : "2016/09/01"[PDat] )) Filters: Publication date to 2016/09/01 | 288 |
| | Search ((((((prognostic model[Title/Abstract] AND ( "0001/01/01"[PDat] : "2016/09/01"[PDat] ))) AND ((validation[Title/Abstract]) OR (verification[Title/Abstract]) OR (test[Title/Abstract]))) AND ( "0001/01/01"[PDat] : "2016/09/01"[PDat] ))) AND external validation[Title/Abstract] Filters: Publication date to 2016/09/01 | 110 |
| diagnostic | Search "diagnostic model"[Title/Abstract] Filters: Publication date to 2016/09/01 | 563 |
| | Search ((((diagnostic model[Title/Abstract] AND ( "0001/01/01"[PDat] : "2016/09/01"[PDat] ))) AND ((validation[Title/Abstract]) OR (verification[Title/Abstract]) OR (test[Title/Abstract])) Filters: Publication date to 2016/09/01 | 208 |
| | Search ((((((((diagnostic model[Title/Abstract] AND ( "0001/01/01"[PDat] : "2016/09/01"[PDat] ))) AND ((validation[Title/Abstract]) OR (verification[Title/Abstract]) OR (test[Title/Abstract]))) AND ( "0001/01/01"[PDat] : "2016/09/01"[PDat] ))) AND (independent[Title/Abstract] OR external[Title/Abstract] ) ( AND ( "0001/01/01"[PDat] : "2016/09/01"[PDat] )) Filters: Publication date to 2016/09/01 | 35 |
| | Search ((((((diagnostic model[Title/Abstract] AND ( "0001/01/01"[PDat] : "2016/09/01"[PDat] ))) AND ((validation[Title/Abstract]) OR (verification[Title/Abstract]) OR (test[Title/Abstract]))) AND ( "0001/01/01"[PDat] : "2016/09/01"[PDat] ))) AND external validation[Title/Abstract] Filters: Publication date to 2016/09/01 | 10 |

Note:  In this paper we focus on prediction models.  While alternative search terms - such as "prognostic tool" or "diagnostic tool" - increase the number of papers, refining the search using terms such as "internal validation" or "external validation" or "validation cohort" or "independent validation" generates a similar pattern of results.  For example a search through to 01 Sept 2016 for "prognostic tool" in the [Title/Abstract] returned 1,715 hits.  Refining the search to papers with the terms "validation" or "confirmation" or "testing" or "checking" reduces the search to just 232 hits.  A specific search for the term "validation" returned just 116 hits: even extending the search to [All Fields] returned just 136 papers.  Narrowing the search to papers including "external validation" returned just 19 papers even after extending the search to [All Fields].

# Building Reality Checks into the Translational Pathway for Diagnostic and Prognostic Models

# Building Reality Checks into the Translational Pathway for Diagnostic and Prognostic Models